\documentclass{jfm}

\usepackage{multiequations}
\usepackage{graphicx,color}
\usepackage{epstopdf}
\usepackage{amsmath,subfig,multiequations}
\usepackage{natbib}

\ifCUPmtlplainloaded \else
  \checkfont{eurm10}
  \iffontfound
    \IfFileExists{upmath.sty}
      {\typeout{^^JFound AMS Euler Roman fonts on the system,
                   using the 'upmath' package.^^J}%
       \usepackage{upmath}}
      {\typeout{^^JFound AMS Euler Roman fonts on the system, but you
                   dont seem to have the}%
       \typeout{'upmath' package installed. JFM.cls can take advantage
                 of these fonts,^^Jif you use 'upmath' package.^^J}%
      }
  \else
  \fi
\fi

\ifCUPmtlplainloaded \else
  \checkfont{msam10}
  \iffontfound
    \IfFileExists{amssymb.sty}
      {\typeout{^^JFound AMS Symbol fonts on the system, using the
                'amssymb' package.^^J}%
       \usepackage{amssymb}%
         
       \let\ge=\geqslant  
      }{}
  \fi
\fi

\ifCUPmtlplainloaded \else
  \IfFileExists{amsbsy.sty}
    {\typeout{^^JFound the 'amsbsy' package on the system, using it.^^J}%
     \usepackage{amsbsy}}
    {}
\fi

\newsavebox{\astrutbox}
\sbox{\astrutbox}{\rule[-5pt]{0pt}{20pt}}

\newcommand{\Ra}{\widetilde{Ra}} 

\newcommand*{\be}{\begin{equation}}
\newcommand*{\ee}{\end{equation}}
\newcommand*{\bea}{\begin{eqnarray}}
\newcommand*{\eea}{\end{eqnarray}}
\newcommand*{\bmeq}{\begin{multiequations}}
\newcommand*{\emeq}{\end{multiequations}}
\newcommand*{\bseq}{\begin{subequations}}
\newcommand*{\eseq}{\end{subequations}}

\def\begineq{\begin{equation}}
\def\endeq{\end{equation}}

\def\begineqn{\begin{equation*}}
\def\endeqn{\end{equation*}}

\def\beginar{\begin{eqnarray}}
\def\endar{\end{eqnarray}}
\def\beginarn{\begin{eqnarray*}}
\def\endarn{\end{eqnarray*}}

\def\lb{\left ( }
\def\rb{\right ) }
\def\lsq{\left [ }
\def\rsq{\right ] }

\def\ep{\epsilon}

\def\Rat{\widetilde{Ra}}

\def\ub{\mathbf{u}}

\def\mub{\overline{\bf u}}

\def\mth{\overline{\vartheta}}
\def\pth{\theta^{\prime}}

\def\pth{\vartheta^{\prime}}

\def\dtau{{\partial_{\tau}}}

\def\dsx{{\partial_x}}
\def\dsy{{\partial_y}}

\def\dst{{\partial_t}}

\def\deta{{\partial_\eta}}
\def\dxi{{\partial_\xi}}

\def\dsz{{\partial_z}}

\def\dz{{\partial_Z}}

\def\hz{{\bf\widehat z}}

\def\litx{{\bf x}}

\def\nabp{{\nabla_{\perp}}}

\def\div{{\nabla \cdot}}
\def\divp{{\nabla_\perp \cdot}}

\def\lp{{\nabla_\perp^2}}

\def\wG{\widetilde{\Gamma}}

\title[Asymptotic equivalence of thermal boundary conditions]{The asymptotic equivalence of fixed heat flux and fixed temperature thermal boundary conditions for rapidly rotating convection}

\author[M. A. Calkins, K. Hale, K. Julien, D. Nieves, D. Driggs and  P. Marti]%
{ Michael A. Calkins$^1$,
Kevin Hale$^2$,
Keith Julien$^1$
David Nieves$^1$,
Derek Driggs$^1$
and Philippe Marti$^1$}

\affiliation{$^1$Department of Applied Mathematics, University of Colorado,
Boulder, CO 80309, USA\\[\affilskip]
$^2$Harvey Mudd College, Claremont, CA 91711, USA}

\pubyear{2010}
\volume{650}
\pagerange{119--126}
\date{?; revised ?; accepted ?. - To be entered by editorial office}

\begin{document}

\maketitle

\begin{abstract}
The influence of fixed temperature and fixed heat flux thermal boundary conditions on rapidly rotating convection in the plane layer geometry is investigated for the case of stress-free mechanical boundary conditions.  It is shown that whereas the leading order system satisfies fixed temperature boundary conditions implicitly, a double boundary layer structure is necessary to satisfy the fixed heat flux thermal boundary conditions.  The boundary layers consist of a classical Ekman layer adjacent to the solid boundaries that adjust viscous stresses to zero, and a layer in thermal wind balance just outside the Ekman layers adjusts the temperature such that the fixed heat flux thermal boundary conditions are satisfied.  The influence of these boundary layers on the interior geostrophically balanced convection is shown to be asymptotically weak, however.  Upon defining a simple rescaling of the thermal variables, the leading order reduced system of governing equations are therefore equivalent for both boundary conditions.  
These results imply that any horizontal thermal variation along the boundaries that varies on the scale of the convection has no leading order influence on the interior convection.
\end{abstract}


\section{\label{sec:INTRO}Introduction}


One of the simplest and most commonly studied systems for investigating convection dynamics is the so-called Rayleigh-B\'{e}nard configuration, consisting of a Boussinesq fluid layer of depth $H$ confined between plane-parallel boundaries, and heated from below.  The constant gravity vector $\mathbf{g} = - g \hz$ points vertically downwards.  Two limiting cases for thermal boundary conditions are often considered when posing the problem mathematically: (1) `perfectly conducting', or fixed temperature (FT), boundary conditions in which the temperature is held fixed along the bounding surfaces; and (2) `perfectly insulating', or fixed flux (FF), boundary conditions in which the normal derivative of the temperature is fixed at the boundaries \citep[e.g.][]{cC80}.  Thermal boundary conditions of geophysical and astrophysical relevance are often considered to reside somewhere between these fixed flux and fixed temperature limits.

For a Newtonian fluid of constant thermal expansivity $\alpha$, kinematic viscosity $\nu$, thermal diffusivity $\kappa$, the non-dimensional Rayleigh number quantifies the strength of the buoyancy force.  For the FT and the FF cases we have, respectively, 
\be
Ra_{FT} = \frac{\alpha g \Delta T H^3}{\nu \kappa}, \quad Ra_{FF} = \frac{\alpha g \beta H^4}{\nu \kappa},
\ee
where $\Delta T$ is the fixed temperature difference between the top and bottom boundaries and $\beta$ is the fixed temperature gradient maintained at the boundaries.
The Prandtl number quantifies the relative importance of viscous and thermal diffusion as $Pr = \nu/\kappa$.  
Upon defining the non-dimensional measure of heat transfer via the Nusselt number,
\be
Nu = \frac{\textnormal{total heat transfer}}{\textnormal{conductive heat transfer}} = \frac{\beta H}{ \Delta T} ,
\ee
it is straightforward to show that the two Rayleigh numbers defined above are related simply by $Ra_{FF} = Nu Ra_{FT}$.  We thus see that for linear convection in which $Nu \equiv 1$ the two Rayleigh numbers are equivalent.  For nonlinear convection in which the critical Rayleigh number has been surpassed, $Nu > 1$ is achieved by adjustment of the temperature gradient $\beta$ at fixed $\Delta T$ for FT boundaries, and vice versa for FF boundaries.

Linear stability shows that for the case of non-rotating convection the most unstable wavenumber is finite for FT boundary conditions \citep[e.g.][]{sC61}, but is zero for FF boundary conditions \citep{dH67}. Although previous work suggests that these differences for linear convection also hold for nonlinear convection \citep{cC80}, numerical simulations of convection show that the statistics for the two cases converge as the Rayleigh number is increased and the flow becomes turbulent \citep{hJ09}.

When the system is rotating with rotation vector $\mathbf{\Omega} =  \Omega \hz$, the Ekman number, $E_H = \nu/2 \Omega H^2 $, is an additional non-dimensional number required to specify the strength of viscous forces relative to the Coriolis force.  The rapidly rotating, quasi-geostrophic convection limit is characterized by $E_H \rightarrow 0$.  As of this writing, only two investigations of FF boundary conditions for the rotating plane layer geometry have been published in the literature, with \cite{tD88} and \cite{sT02} examining the weakly rotating and rapidly rotating linear cases, respectively.  \cite{sT02} utilized a modal truncation approach to show that the critical parameters for the two cases should converge as $E_H\rightarrow 0$; the present work confirms this suggestion.  


In the present work we distinguish between `interior' and 'boundary layer' dynamics, and show that the interior governing equations are identical for the two different thermal boundary conditions upon a simple rescaling of the Rayleigh number and temperature.  Because the $E_H\rightarrow 0$ limit is a singular perturbation of the governing equations, the interior equations cannot satisfy the FF boundary conditions at leading order; a double boundary layer structure is necessary to adjust both the horizontal viscous stresses and the normal derivative of the temperature fluctuation to zero \cite[c.f.][]{wH71}.  It is shown that the boundary layer corrections are asymptotically weak, however, showing that to leading order the interior quasi-geostrophic convection dynamics are equivalent for both thermal boundary conditions.

In section \ref{S:Lin} we present the linear stability of the full Boussinesq Navier-Stokes equations.  In section \ref{S:Asymp} we present the asymptotic reduction of the Navier-Stokes equations in the rapidly rotating limit and concluding remarks are given in section \ref{S:discuss}.

\section{Linear stability of the Navier-Stokes equations}
\label{S:Lin}

In the present section we briefly present the linear stability of the Boussinesq Navier-Stokes equations for both FT and FF thermal boundary conditions.  Upon scaling lengths with the depth of the fluid layer $H$ and time with the viscous diffusion time $H^2/\nu$, the linear system becomes
\be
  \dst \ub + \frac{1}{E_H} \, \hz \times \ub = - \frac{1}{E_H} \nabla p +  \frac{Ra}{Pr} \, \pth \, \hz + \nabla^2 \ub, \label{E:momlin} 
\ee
\be
\dst \pth - w =  \frac{1}{Pr} \nabla^2 \pth, \label{E:heatlin} 
\ee
\be
\div \ub =  0 , \label{E:contlin}
\ee
where the velocity vector is denoted by $\ub=(u,v,w)$, and the temperature is decomposed into mean and fluctuating variables according to $\vartheta = \mth + \pth$.  For both sets of thermal boundary conditions $\mth = 1-z$ and the fluctuating thermal boundary conditions therefore become
\be
\pth = 0, \quad \textnormal{at} \quad z = 0, 1, \quad (FT) \label{E:FT2} 
\ee
\be
\dsz \pth = 0, \quad \textnormal{at} \quad z = 0, 1 . \quad (FF) \label{E:FF2}
\ee
Stress-free, impenetrable mechanical boundary conditions on the top and bottom boundaries are assumed throughout and given by
\be
w = \dsz u = \dsz v = 0, \quad \textnormal{at} \quad z = 0, 1.
\ee

The system \eqref{E:momlin}-\eqref{E:contlin} is discretized in the vertical and horizontal dimensions with Chebyshev polynomials and Fourier modes respectively, and formulated as a generalized eigenvalue problem.  We solve the system in primitive variable form and enforce boundary conditions via the tau method.  The eigenvalue problem is solved with Matlab's `sptarn' function.  For further details of the numerical methods the reader is referred to \cite{mC13}.

\begin{figure}
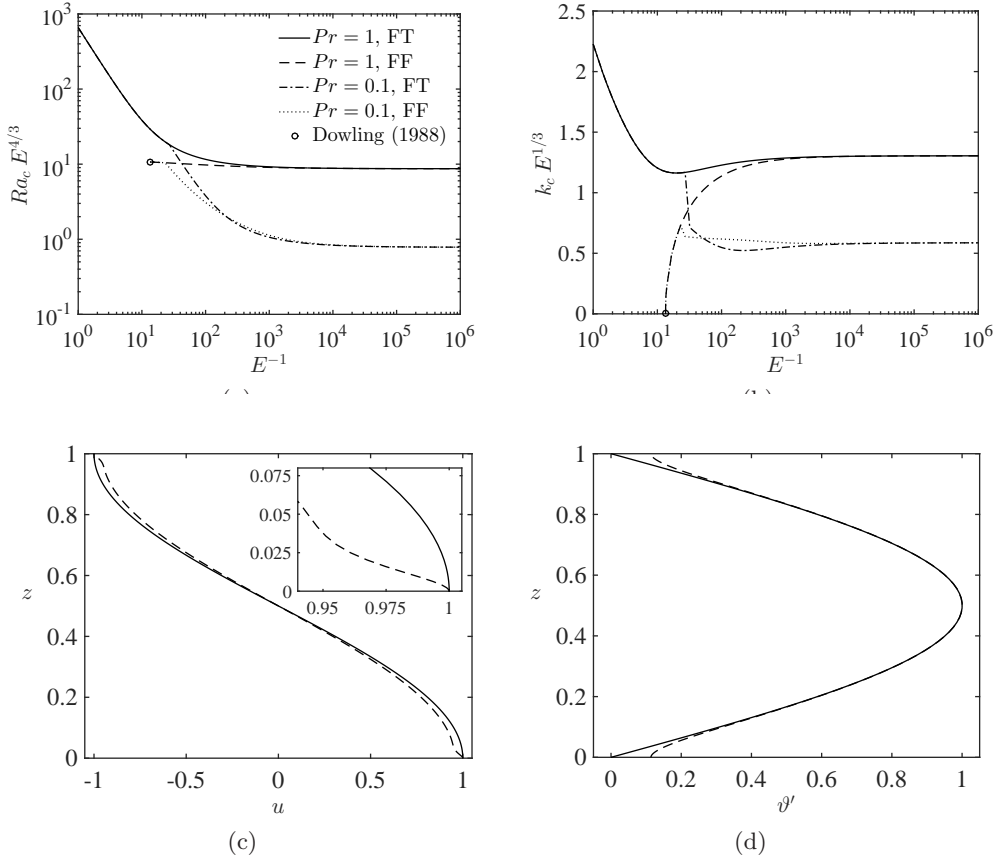

  \begin{center}
   \subfloat[]{
      \includegraphics[height=5cm]{Rc_vs_E}}
      \qquad
    \subfloat[]{
      \includegraphics[height=5cm]{kc_vs_E}} \\
   \subfloat[]{
      \includegraphics[height=5cm]{u_eigTa1e8P1}}
      \qquad
    \subfloat[]{
      \includegraphics[height=5cm]{T_eigTa1e8P1}}      
  \end{center}
\caption{Linear stability of the Navier-Stokes equations for fixed temperature (FT) and fixed flux (FF) thermal boundary conditions.  (a) Asymptotically scaled critical Rayleigh number and (b) critical wavenumber as a function of the inverse Ekman number for both steady ($Pr=1$) and oscillatory ($Pr=0.1$) convection.  (c) Horizontal velocity and (d) temperature eigenfunctions for $Pr=1$ and $E_H=10^{-4}$; the inset figure in (c) shows a magnified view of the Ekman layer along the bottom boundary.  In (a) and (b) the Ekman number of $E_H=0.0745$ at which the critical wavenumber for FF boundary conditions becomes non-zero is shown by the open circle, as first determined by \cite{tD88}.}
\label{F:linNS}
\end{figure}

Figure \ref{F:linNS} shows results from the linear stability calculations. Results are given for both steady ($Pr=1$) and oscillatory ($Pr=0.1$) convection; we note oscillatory convection does not exist for $Pr\ge1$ and becomes the primary instability for $Pr\lesssim 0.68$ \citep{sC61}.  For $E_H \lesssim 10^{-5}$, both the asymptotically scaled critical Rayleigh number $Ra_c E^{4/3}$ (Figure \ref{F:linNS}a)  and wavenumber $k_c E^{1/3}$ (Figure \ref{F:linNS}b) obtained from FT and FF boundary conditions are observed to converge to nearly equivalent values.  The open circle shows the Ekman number $E_H=0.0745$ calculated by \cite{tD88} at which the instability becomes characterized by $k_c \ne 0$ with $Ra_c = 341.05$.


Figure \ref{F:linNS}c shows the horizontal ($x$) velocity eigenfunction for $E_H=10^{-4}$ where Ekman layers can be seen at the top and bottom boundaries for the FF (dashed curve) case; a magnified view of the bottom Ekman layer is shown in the inset figure.  The temperature perturbation eigenfunctions shown in Figure \ref{F:linNS}d show that both the FT and FF cases have identical structure in the fluid interior, whereas a thermal boundary layer is observed in the FF case.  In the following section we present the asymptotic reduction of the Navier-Stokes equations to better understand and quantify this behavior.

\section{Asymptotics}
\label{S:Asymp}

To proceed with the asymptotic development, we follow the work of \cite{mS06} and write the governing equations using a generic non-dimensionalization such that
\begin{gather}
D_t \ub  + \frac{1}{Ro} \hz \times \ub   =  - Eu \nabla p + \Gamma \theta \hz  + \frac{1}{ Re} \nabla^2 \ub,  \label{E:mom1}  \\
\nabla \cdot \ub  =  0, \label{E:mass1}  \\
D_t \vartheta =  \frac{1}{Pr Re} \nabla^2 \vartheta ,  \label{E:energy1}
\end{gather}
where $D_t (\cdot) = \dst(\cdot) + \ub \cdot \nabla(\cdot)$ and the velocity, pressure and temperature are denoted by $\ub$, $p$, and $\vartheta$, respectively.  The above system has been non-dimensionalized utilizing the velocity scale $U$, length $L$, time $L/U$, pressure $P$, and temperature $\widetilde{T}$. For the FT and FF cases the temperature scale becomes $\Delta T$ and 
$Nu  \Delta T$,
respectively.  The Rossby, Euler, buoyancy and Reynolds numbers are defined by
\be
Ro = \frac{U}{2 \Omega L}, \quad 
Eu = \frac{P}{\rho_0 U^2}, \quad \Gamma = \frac{g \alpha \widetilde{T}}{U^2}, \quad Re = \frac{U L}{\nu} .
\ee

In the present work we are interested in the $\ep \equiv Ro \rightarrow 0$ limit. In the fluid interior we employ multiple scales in the axial space direction and time such that
\be
\dsz \rightarrow \dsz + \ep \dz , \quad
\dst \rightarrow \dst + \ep^2 \dtau ,
\ee
where $Z = \ep z$ is the large-scale vertical coordinate and $\tau = \ep^2 t$ is the `slow' timescale.  It has been shown that the following distinguished limits can be taken to reduce the governing equations to accurately model quasi-geostrophic convection \citep[e.g.][]{mS06}
\be
Eu = \frac{1}{\ep^2}, \quad \Gamma = \frac{\wG}{\ep}, \quad Re  = O(1) , \quad Pr = O(1),
\ee
where $\wG=O(1)$.  Scaling the velocity viscously such that $U = \nu/L$ we have 
\be
\ep = E^{1/3}, \quad \wG = \frac{E^{4/3} Ra}{Pr}, \quad Re = 1 ,
\ee
where the $L$-scale Ekman number is related to the $H$-scale Ekman number via $E=E_H\ep^{-2}$, i.e.~$L=E^{1/3} H$.  We keep the notation for the Rayleigh number generic in the sense that $Ra$ denotes either $Ra_{FT}$ or $Ra_{FF}$ depending upon the particular boundary conditions employed.  Hereafter, we define the asymptotically reduced Rayleigh number as $\Rat \equiv E^{4/3} Ra$.  

We utilize a composite asymptotic expansion approach \citep[e.g.][]{aN08} and, following \cite{wH71}, decompose each variable into interior $(i)$, middle $(m)$, and Ekman layer $(e)$ components.  For instance, the dependent variable $f$ can be written as
\be
f= f^{(i)}(\litx,Z,t,\tau) + f^{(m)}(x,y,\xi,t) + f^{(e)}(x,y,\eta,t) , \label{E:decomp}
\ee
where $\xi=z$ and $\eta = \ep^{-1/2}z$ are boundary layer variables.  The above representation ensures that each dependent variable is uniformly valid throughout the domain.  The boundary layer variables consist of a sum of contributions from the top and bottom boundary layers; for brevity, we focus on the bottom boundary layers. In the present work we make use of the following limits and notation 
\be
\lim_{\xi \rightarrow \infty} \lb f^{(m)} \rb = \lb f^{(m)} \rb^{(i)} = \lim_{\eta \rightarrow \infty}  \lb f^{(e)} \rb = \lb f^{(e)} \rb^{(i)} = 0 , \label{E:limits1}
\ee
\be
\lim_{\xi \rightarrow 0} \lb f^{(i)} \rb = \lim_{\eta \rightarrow 0} \lb f^{(i)} \rb = f^{(i)}\lb Z = 0\rb = f^{(i)}(0) . \label{E:limits2}
\ee
We then expand each variable in a power series according to
\be
f^{(i)}(\litx,Z,t)  = f_0^{(i)}(\litx,Z,t,\tau) + \ep^{1/2} f_{1/2}^{(i)}(\litx,Z,t,\tau) + \ep f_{1}^{(i)}(\litx,Z,t,\tau) + O(\ep^{3/2}) . \label{E:expand}
\ee
Each dependent variable is further decomposed into mean and fluctuating components such that
\be
f^{(i)}(\litx,Z,t,\tau) = \overline{f}^{(i)}(Z,\tau) + f^{\prime (i)}(\litx,Z,t,\tau) ,
\ee
where the averaging operator is defined by
\be
\overline{f}(Z,\tau) = \lim_{[\tau], [A] \rightarrow \infty} \, \frac{1}{[\tau] [A]} \int_{[\tau],[A]} f \, dx dy , \qquad\mbox{and}\qquad \overline{f'} \equiv 0.
\ee

\subsection{The interior equations}

By substituting decompositions for each variable of the form \eqref{E:decomp} into the governing equations and utilizing the limits \eqref{E:limits1}-\eqref{E:limits2}, equations for each region can be derived; expansions of the form \eqref{E:expand} are then utilized to determine the asymptotic behavior of each fluid region.  Because the derivation of the interior equations has been given many times previously, we present only the salient features and direct the reader to previous work \citep[e.g.][]{mS06} for details on their derivation.  The main point is that the interior convection is geostrophically balanced and horizontally divergence-free to leading order
\be
\hz \times \ub^{(i)}_{0} = - \nabla_\perp p^{(i)}_{1}, \quad \divp \ub^{(i)}_{0,\perp} = 0,
\ee
where $\nabla_{\perp} = (\dsx,\dsy,0)$.  The above relations allow us to represent the geostrophic velocity via the geostrophic streamfunction $\psi^{(i)}_0 \equiv p^{\prime(i)}_1$ such that  $\ub^{(i)}_{0,\perp} = -\nabla \times \psi^{(i)}_0\hz$.  The vertical vorticity is then $\zeta^{(i)}_{0} = \nabla_\perp^2 \psi^{(i)}_0$.  The interior vertical vorticity, vertical momentum, fluctuating heat, and mean heat equations then become  
\begin{gather}
  D^{\perp}_{t} \zeta^{(i)}_{0}  - \dz w^{\prime(i)}_0 = \lp \zeta^{(i)}_{0}, \label{E:vortint} \\
  D^{\perp}_{t} w^{\prime(i)}_0 + \dz \psi^{(i)}_0   = \frac{\Rat}{Pr} \vartheta_1^{\prime(i)}  + \lp w^{\prime(i)}_0, \label{E:momint} \\
  D^{\perp}_{t}\vartheta_1^{\prime(i)}  + w^{\prime(i)}_0 \dz \overline{\vartheta}^{(i)}_0  =  \frac{1}{Pr} \lp   \vartheta_1^{\prime(i)}, \label{E:heatint} \\
  \dtau \overline{\vartheta}_0^{(i)} + \dz \overline{\lb w^{\prime(i)}_0 \vartheta_1^{\prime(i)} \rb} = \frac{1}{Pr} \partial^2_{Z} \overline{\vartheta}_0^{(i)}, \label{E:mheatint}
\end{gather}
where $D_t^{\perp} (\cdot) = \dst(\cdot) + \ub \cdot \nabla_{\perp}(\cdot)$ .  The mean interior velocity field $\mub_0^{(i)}$ is zero and the mean momentum equation reduces to hydrostatic balance in the vertical, $\dz \overline{p}^{(i)}_0 = (\Rat/Pr) \overline{\vartheta}_0^{(i)}$.  

The interior system is fourth order with respect to the large-scale vertical coordinate $Z$.  Two boundary conditions are supplied by impenetrability such that $w^{\prime(i)}_0(0) =  w^{\prime(i)}_0(1) = 0$.  Although no $Z$-derivatives with respect to $\vartheta^{\prime(i)}_1$ are present in equation \eqref{E:heatint}, evaluating this equation at the boundaries shows the FT conditions $\vartheta^{\prime(i)}_1(0)= \vartheta^{\prime(i)}_1(1)=0$ are satisfied implicitly for the fluctuating temperature. Evaluating equation \eqref{E:momint} at either the top or bottom boundary with the use of impenetrability shows that stress-free boundary conditions are implicitly satisfied as well since 
$\dz \psi^{(i)}_0(0) = \dz \psi^{(i)}_0(1) = 0$.  

For the case of FT thermal boundary conditions, we have 
\be
\overline{\vartheta}_0^{(i)}(0) = 1, \quad \textnormal{and}  \quad \overline{\vartheta}_0^{(i)}(1) = 0.  \quad (FT)
\ee
Thus, for the FT case the boundary layer corrections are identically zero and the above system is complete.  Numerous investigations have used the above system of equations to investigate rapidly rotating convection in the presence of stress-free mechanical boundary conditions and have shown excellent agreement with direct numerical simulations (DNS) of the Navier-Stokes equations and laboratory experiments \citep{sS14,jmA15}.  

For the FF case the mean temperature boundary conditions become 
\be
\dz \overline{\vartheta}_0^{(i)}(0) = -1, \quad \textnormal{and}  \quad \dz \overline{\vartheta}_0^{(i)}(1) = -1. \quad (FF)
\ee  
We further require $\dz \vartheta_1^{\prime(i)}(0) = \dz \vartheta_1^{\prime(i)}(1)= 0$; boundary layer corrections are therefore required since these conditions are not satisfied by \eqref{E:heatint}.  In the following two subsections we determine the magnitude of these boundary layer corrections.


\subsection{The middle layer equations}

The first non-trivial fluctuating middle layer momentum equation occurs at $O(\ep)$ to yield the thermal wind balance
\be
\hz \times \ub^{\prime (m)}_2 = - \nabla p^{\prime (m)}_{3} + \frac{\Rat}{Pr} \vartheta^{\prime(m)}_2 \hz , \label{E:mommid2}
\ee 
such that $\divp \ub^{\prime (m)}_{2} = 0$ and $w^{\prime (m)}_{2} \equiv 0$.  
The mean velocity field $\mub^{(m)} \equiv 0$.  

The leading order temperature equation for the middle layer is
 \be
\dst \vartheta^{\prime(m)}_2 + \ub^{\prime (i)}_0 (0) \cdot \nabp \vartheta^{\prime(m)}_2  = \frac{1}{Pr} \nabla^2 \vartheta^{\prime(m)}_2 , 
\ee
with corresponding boundary conditions
\be
\dz \vartheta^{\prime(i)}_{1}(0) + \dxi \vartheta^{\prime(m)}_{2}(0) = 0, \quad \vartheta^{\prime(m)}_{2}(\xi \rightarrow \infty) \rightarrow 0 . \label{E:midthermBC}
\ee
We find the first non-trivial mean temperature to be of magnitude $O(\ep^5)$ and therefore omit any further consideration of this correction.

The first three orders of the stress-free mechanical boundary conditions along the bottom boundary become
\be
\dz \ub^{\prime (i)}_{0,\perp}(0) = 0, \quad  
\dz \ub^{\prime (i)}_{1/2,\perp}(0) = 0, \quad 
\dz \ub^{\prime (i)}_{1,\perp}(0) + \dxi \ub^{\prime (m)}_{2,\perp}(0) + \deta \widetilde{\ub}^{\prime (e)}_0(0) = 0 . \label{E:stressBC}
\ee
Thus, the first two orders of the interior velocity satisfy stress free conditions on their own and therefore need no boundary layer correction.  Here we have rescaled the Ekman layer velocity according to $\ub^{\prime (e)}_{5/2} = \ep^{5/2} \widetilde{\ub}^{\prime (e)}_0$; this rescaling is simply highlighting the fact that the Ekman layer velocities are significantly weaker than those in the interior.

\subsection{The Ekman layer equations}

The Ekman layer equations have been studied in great detail in previous work \citep[e.g.][]{hG68}, so we simply state the leading order continuity and momentum equations as
\be
\divp \widetilde{\ub}^{\prime (e)}_0 + \partial_{\eta} \widetilde{w}^{\prime (e)}_{\frac{1}{2}} = 0, \quad \hz \times \widetilde{\ub}^{\prime (e)}_0 = \partial^2_{\eta} \widetilde{\ub}^{\prime (e)}_0
\ee
where $w_3^{\prime(e)}=\epsilon^3  \widetilde{w}^{\prime (e)}_{\frac{1}{2}} $. All of the mean Ekman layer variables can be shown to be zero.  A key component in the present analysis that differs from previous work is the middle, thermal wind layer that enters the Ekman layer solution via the stress-free boundary conditions \eqref{E:stressBC}.  Utilizing the thermal wind relations for the middle layer that follow from equation \eqref{E:mommid2},
\be
\dxi u^{\prime (m)}_2 = -\frac{\Rat}{Pr} \dsy \vartheta^{\prime(m)}_2, \quad \dxi v^{\prime (m)}_2 = \frac{\Rat}{Pr} \dsx \vartheta^{\prime(m)}_2 ,
\ee
the stress-free boundary conditions along the bottom boundary can be written as
\be
\dz \mathbf{u}^{\prime (i)}_{1,\perp}(0) + \frac{\Rat}{Pr} \nabla^{\perp} \vartheta^{\prime(m)}_2(0) + \deta \widetilde{\mathbf{u}}^{\prime (e)}_0(0) = 0 ,
\ee
where $\nabla^{\perp} = (-\dsy,\dsx,0)$.  Solving the Ekman layer momentum equations for the horizontal components of the velocity field with the additional requirement that $(\widetilde{u}^{\prime (e)}_0,\widetilde{v}^{\prime (e)}_0) \rightarrow 0$ as $\xi \rightarrow \infty$, the continuity equation is then used to find the Ekman pumping velocity 
\be
 \widetilde{w}^{\prime (e)}_{\frac{1}{2}} = \lsq \dz \zeta^{(i)}_1(0) + \frac{\Rat}{Pr} \lp \vartheta^{\prime(m)}_2(0) \rsq e^{-\frac{\eta}{\sqrt{2}}} \cos \lb \frac{\eta}{\sqrt{2}} \rb . \label{E:ekpump}
\ee
Thus, vertical velocities of magnitude $O(\ep^{3})$ are induced by FF thermal boundary conditions and result from both finite viscous stresses within the fluid interior and horizontal variations of the temperature within the middle layer.  This finding is closely analogous to the Ekman pumping effect first reported by \cite{rH64} for shallow layer quasi-geostrophic flow in the presence of lateral temperature variations along a free surface.  Evaluating equation \eqref{E:ekpump} at $\eta=0$ provides a parameterized boundary condition for the effects of Ekman pumping.

The small magnitude of the Ekman pumping velocity \eqref{E:ekpump} results in very weak $O(\ep^5)$ temperature fluctuations within the Ekman layer.  Because of this, the dominant correction of the FF thermal boundary conditions occurs within the middle layer and we do not consider the Ekman layer temperature any further.


\subsection{Synthesis}

The thermal boundary layer correction given by equation \eqref{E:midthermBC} is passive in the sense that $\vartheta^{\prime(m)}_2$ can be calculated a posteriori with knowledge of $\vartheta^{\prime(i)}_1$.  Thus, the leading order interior dynamics are insensitive to the thermal boundary conditions.  The Ekman layer analysis shows that the first six orders of the interior vertical velocity satisfy the impenetrable mechanical boundary conditions $w^{\prime(i)}_i(0) = 0$, for $i=0,\ldots, 5/2$.
At $O(\ep^3)$ we have the Ekman pumping boundary conditions
\be
w^{\prime(i)}_3(0) = - w^{\prime(m)}_3(0) - \dz \zeta^{(i)}_1(0)  - \frac{\Rat}{Pr} \lp \vartheta^{\prime(m)}_2(0) ,
\ee
where we have used the Ekman pumping relation \eqref{E:ekpump} evaluated at $\eta=0$.  From the standpoint of linear theory, the first correction to the critical Rayleigh number will therefore occur at $O(\ep^3)$; this explains the linear behavior previously discussed in section \ref{S:Lin}.

\section{Discussion}
\label{S:discuss}

In light of the boundary layer analysis, we conclude that the leading order quasi-geostrophic dynamics are described by equations \eqref{E:vortint}-\eqref{E:mheatint} for \textit{both} FT and FF thermal boundary conditions.  Indeed, inspection of the system shows that it is invariant under the following rescaling of the Rayleigh numbers and temperature variables,
\be
\Rat_{FT} = \frac{\Rat_{FF}}{Nu}, \quad \vartheta_{1,FT}^{\prime(i)} = Nu \, \vartheta_{1,FF}^{\prime(i)}, \quad \mth_{0,FT}^{(i)} = Nu \, \mth_{0,FF}^{(i)} . \label{E:scaling}
\ee
Integrating the time-averaged mean heat equation with respect to $Z$ yields
\be
Pr \overline{\lb w^{\prime(i)}_0 \vartheta_{1,FT}^{\prime(i)} \rb} = \dz \mth_{0,FT}^{(i)} + Nu, \quad (FT) \label{E:mFT}
\ee
\be
Pr \overline{\lb w^{\prime(i)}_0 \vartheta_{1,FF}^{\prime(i)} \rb} = \dz \mth_{0,FF}^{(i)} + 1, \quad (FF) \label{E:mFF}
\ee
for the FT and FF cases, respectively.  The appropriate thermal boundary conditions have been applied at $Z=0$ in the above relations.  Taking either equation \eqref{E:mFT} or \eqref{E:mFF} and utilizing \eqref{E:scaling} shows that the mean interior temperature gradient is described by identical equations for the two cases.  This leading order correspondence is the result of the anisotropic spatial structure of rapidly rotating convection. 

\begin{figure}
  \begin{center}
    \subfloat[]{
      \includegraphics[height=5cm]{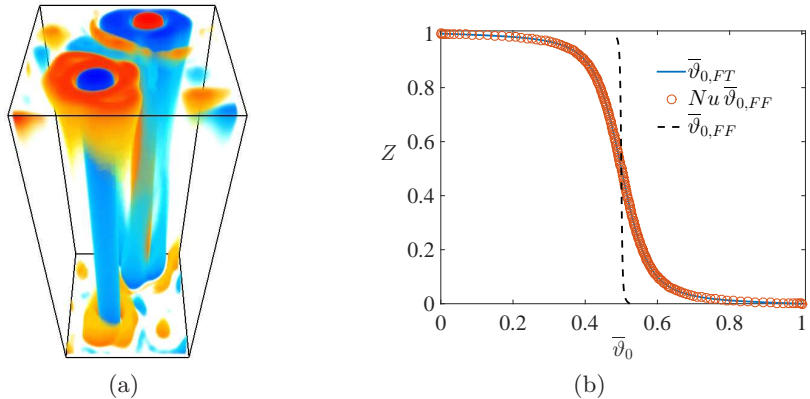}}
      \hspace{1.5cm}
    \subfloat[]{
      \includegraphics[height=4.5cm]{theta_bar_profiles}}
  \end{center}
 \caption{(a) An example volumetric rendering of the temperature perturbation from a simulation of the quasi-geostrophic convection equations showing the `convective Taylor column' (CTC) regime.  (b) Mean temperature profiles obtained with both FT (solid blue) and FF (dashed black) boundary conditions, and the rescaled FF temperature profile (red open circles). The parameters are $\Pr=7$, $\Ra_{FT} = 46.74$, $\Ra_{FF} = 1000$, and $Nu=21.39$.  
 }
 \label{F:nonlin}
\end{figure}

The above results indicate that the findings of previous work on low Rossby number convection employing FT thermal boundary conditions can be accurately applied to the case of FF thermal boundary conditions by use of the rescalings given by equations \eqref{E:scaling}.  \cite{kJ12} identified four flow regimes that occur in rapidly rotating convection as a function of the Prandtl and (FT) Rayleigh numbers.  The so-called `convective Taylor column' (CTC) regime is distinguished by coherent, vertically aligned convective structures that span the depth of the fluid.  Figure \ref{F:nonlin}(a) shows a volumetric rendering of the temperature perturbation for $Pr=7$ and $\Rat_{FT}=46.74$, or $\Rat_{FF}=1000$ and $Nu=21.39$. The CTC regime occurs over the FT Rayleigh number range of  $20\lesssim \Rat_{FT} \lesssim 55$, corresponding to a FF Rayleigh number range of $82\lesssim \Rat_{FF} \lesssim 1656$ \citep[e.g.~see][]{dN14}.   Figure \ref{F:nonlin}(b)  shows mean temperature profiles obtained utilizing the FT and FF thermal boundary conditions, along with the remapped FF mean temperature profile.  The Nusselt number $Nu=21.39$ corresponds to a mean temperature difference of $0.0468$ between the top and bottom boundaries for the FF case.

Of particular interest in convection studies is the dependence of the heat transfer scaling with the strength of the thermal forcing input via Nusselt-Rayleigh number scalings of the form $Nu \sim \Rat_{FT}^\alpha$.  With the rescaling given in \eqref{E:scaling} the FF equivalent of this relation becomes $Nu \sim \Rat_{FF}^\beta$ where $\beta  = \alpha/(\alpha+1)$.  For the CTC regime the exponent is $\alpha \approx 2.1$ \citep{kJ12}, yielding $\beta \approx 0.68$.  Additionally, the final regime of geostrophic turbulence achieves a dissipation-free scaling law with $\alpha = 3/2$ such that $\beta = 3/5$ \citep{kJ12b}. Similarly, the dependence of all other variables of interest on the Rayleigh number (e.g. mean temperature gradient, vorticity, etc.) can also be remapped to the case of FF thermal boundary conditions.

\section{Conclusion}
\label{S:conclude}

In the present work we have shown that the leading order dynamics of rapidly rotating convection in a plane layer geometry are equivalent for both FT and FF thermal boundary conditions.  FF thermal boundary conditions give rise to a double boundary layer structure in the limit of rapid rotation that is asymptotically weak.  Our findings suggest that all previous work employing FT thermal boundary conditions also accurately describes FF thermal boundary conditions within the regime of asymptotic validity, i.e. $Ro\sim E^{1/3}\ll1$ and $\widetilde{Ra} \lesssim \mathcal{O} (E^{-1/3})$ \citep{kJ12b}.  This result adds to the robustness of the reduced, quasi-geostrophic model in geophysical  and astrophysical applications where stress-free boundary conditions are typically assumed; recent work has also extended the model to the case of no-slip boundary conditions \citep{kJ15} that are of relevance for laboratory experiments and planetary interiors \citep[e.g.][]{jmA15,jC15}.


An interesting consequence of these findings is that any horizontal thermal variation along the boundaries that varies on the scale of the convection has no leading order influence on the interior convection.  This finding helps to explain the results of previous spherical convection studies investigating the role of temperature variations along the outer boundary \citep[e.g.][]{cD09}.  




\bibliographystyle{jfm}

\bibliography{Biblio.bib}

\end{document}